\documentclass[letter,10pt,conference,twoside]{IEEEtran}

\usepackage{caption}

\DeclareCaptionLabelFormat{lc}{{#1}~#2}
\def\tablename{Table}

\newenvironment{rrcases}
  {\left.\begin{aligned}}
  {\end{aligned}\right\rbrace}

\ifCLASSINFOpdf
\else
\fi
\usepackage{url}
 \usepackage{epsfig,endnotes}
\let\labelindent\relax
\usepackage{floatpag,enumitem}
\usepackage{hyphenat}

\usepackage{cite}
\usepackage{relsize}
  \usepackage[top=.75in, bottom=1in, left=1in, right=1in]{geometry}
 \usepackage{tikz}

\newcommand{\de}{\stackrel{\text{def}}{=}}
\newcommand{\con}{\boldsymbol{\mid}}

\usepackage{multirow}
\usepackage{setspace}
\usepackage{mathrsfs}
\usepackage{bbm}
\usepackage{pifont}
\usepackage{amsfonts}

\usepackage{amsmath, amsthm}
\usepackage{tabularx}
\usepackage{listings}
\usepackage{graphicx}
\usepackage{amssymb}
\usepackage{array}
\usepackage{cases}

 \usepackage{supertabular}
\usepackage{mathrsfs}

\usepackage{psfrag}
\usepackage{bbding}
\newcommand{\bcap} {\hspace{2pt} \mathlarger{\cap}
\hspace{2pt}}
\newcommand{\bcup} {\hspace{2pt} \mathlarger{\cup}
\hspace{2pt}}

\usepackage{float}

\usepackage{amsbsy}

\makeatletter
\providecommand{\leftsquigarrow}{%
  \mathrel{\mathpalette\reflect@squig\relax}%
}
\newcommand{\reflect@squig}[2]{%
  \reflectbox{$\m@th#1\rightsquigarrow$}%
}
\makeatother
\usepackage{algorithm}
 \usepackage{algorithmic}

\makeatletter
\newcommand{\newalgname}[1]{%
  \renewcommand{\ALG@name}{#1}%
}

\newcommand {\C} {{\rm I\kern-5.5pt C}}

\newcommand{\bp}[1]{{\mathbb{P}}\left[{#1}\right]}
\newcommand{\bigp}[1]{{\mathbb{P}}\big[{#1}\big]}

\newcommand{\bfu}[1]{{\mathbb{F}}\left[{#1}\right]}

\newcommand{\fsquare}{\vrule height6pt width7pt depth1pt}   % filled square
\def\centerhack#1{\hbox to 0pt{\hss\footnotesize #1\hss}}
\def\centerhackn#1{\hbox to 0pt{\hss #1\hss}}
\def\dchack#1{\vbox to 0pt{\vss{\hbox to 0pt{\hss#1\hss}}\vss}}
\newcommand{\pr}{\mathbb{P}} % probability measure

\usepackage{etoolbox}
\AtBeginEnvironment{align}{\setcounter{subeqn}{0}}% Reset subequation number at start of align
\newcounter{subeqn} %
\usetikzlibrary{positioning}

\newcounter{mysub}

\newtheorem{thm}{Theorem}

\newtheorem*{proposition1.1}{Proposition 1.1}
\newtheorem*{proposition1.2}{Proposition 1.2}
\newtheorem*{proposition1.3}{Proposition 1.3}
\newtheorem*{proposition2.1}{Proposition 2.1}
\newtheorem*{proposition2.2}{Proposition 2.2}
\usepackage{subfigure}
\IEEEoverridecommandlockouts
\begin{document}

\title{Composition Properties of Bayesian Differential Privacy}

% \author{~\\Jun~Zhao,~\IEEEmembership{Member,~IEEE}\\{\color{blue}Almost all parts of the paper have been improved, so we do not highlight the changes in   color\\for better readability. Instead, we explain the changes in the revision letter.}\vspace{-35pt}\\~\thanks{The author is with the Cybersecurity Lab (CyLab) and Department of Electrical and Computer Engineering,
% Carnegie Mellon University, Pittsburgh, PA 15213. Email: junzhao@cmu.edu}}
%\date{}
% \author{Jun~Zhao,~\IEEEmembership{Member,~IEEE}\thanks{The author was with the Cybersecurity Lab (CyLab) at Carnegie Mellon University, Pittsburgh, PA 15213, USA. He is now with Arizona State University, Tempe, AZ 85281, USA, and Princeton University, Princeton, NJ 07450, USA (Email: junzhao@alumni.cmu.edu).}}

\author{\IEEEauthorblockN{Jun Zhao\thanks{The author Jun Zhao obtained his PhD from Carnegie Mellon University, Pittsburgh, PA 15213, USA, where he was with the Cybersecurity Lab (CyLab). He was a postdoctoral scholar with Arizona State University, Tempe, AZ 85281, USA. He is now a research fellow at Nanyang Technological University in Singapore.
 Email: \texttt{junzhao@alumni.cmu.edu} \newline \indent \textit{\textbf{2017 IEEE 28th Annual International Symposium on Personal, Indoor, and Mobile Radio Communications (PIMRC):}} Special Session SP-04 on ``Resource-Efficient, Reliable and Secure Internet of Things in the 5G Era'' \newline \indent 978-1-5386-3531-5/17/\$31.00~\copyright~2017 IEEE}}
\IEEEauthorblockA{\texttt{junzhao@alumni.cmu.edu}\\[0pt]~}}

\maketitle

%\pagestyle{plain}

%\thispagestyle{plain}

%\author{{Jun Zhao}
%CyLab and Dept.
%of ECE \\
%Carnegie Mellon University \\
%Pittsburgh, PA 15213\\
%Email: junzhao@cmu.edu \and {Osman Ya\u{g}an}
%CyLab and Dept.
%of ECE\\
%Carnegie Mellon University \\
%Moffett Field, CA 94035\\
%Email: oyagan@ece.cmu.edu \and {Virgil Gligor}
%CyLab and Dept.
%of ECE \\
%Carnegie Mellon University \\
%Pittsburgh, PA 15213\\
%Email: gligor@cmu.edu}

%
%\author{\IEEEauthorblockN{Jun Zhao, Osman Ya\u{g}an and Virgil Gligor}
%\IEEEauthorblockA{\\CyLab and Dept.
%of ECE \\
%Carnegie Mellon University \\ \{junzhao, oyagan,
%virgil\}@andrew.cmu.edu}}

%\author{\IEEEauthorblockN{Jun Zhao, Osman Ya\u{g}an and Virgil Gligor}
%\IEEEauthorblockA{{\tiny~\vspace{-7pt}}\\CyLab and Dept.
%of ECE \\
%Carnegie Mellon University \\ \{junzhao, oyagan,
%virgil\}@andrew.cmu.edu\vspace{-7pt}}}

%\author{\IEEEauthorblockN{~}
%\IEEEauthorblockA{~\\
%~ \\ ~}}

%\markboth{} {\large This paper has been accepted at IEEE ISIT 2014
%to be held in June--July 2014.}

\begin{abstract}

Differential privacy is a rigorous privacy standard that has been
applied to a range of data analysis tasks. To broaden the application scenarios of differential privacy when data records have dependencies, the notion of Bayesian differential privacy has been recently proposed. However, it is unknown whether Bayesian differential privacy   preserves three nice properties of differential privacy: \emph{sequential composability}, \emph{parallel composability}, and \mbox{\emph{post-processing}}. In this paper, we provide an affirmative answer to this question; i.e., Bayesian differential privacy still have these properties.
The idea behind \emph{sequential composability} is that if we have $m$ algorithms $Y_1, Y_2, \ldots, Y_m$,
where $Y_{\ell}$
is independently \mbox{$\epsilon_{\ell}$-Bayesian differential private} for ${\ell}=1,2,\ldots,m$, then by feeding the
result of $Y_1$ into $Y_2$, the
result of $Y_2$ into $Y_3$, and so on, we will finally have an
\mbox{$\sum_{\ell=1}^m \epsilon_{\ell}$-Bayesian differential private} algorithm. For \emph{parallel composability}, we consider the situation where a   database is partitioned into $m$ disjoint subsets. The \mbox{$\ell$-th}  subset is input to a Bayesian differential private algorithm $Y_{\ell}$, for ${\ell}=1,2,\ldots,m$. Then the parallel composition of $Y_1$,  $Y_2$, $\ldots$,  $Y_m$ will be \mbox{$\max_{\ell=1}^m \epsilon_{\ell}$-Bayesian differential private}. The \emph{post-processing} property means that a data analyst,
without additional knowledge about the private database, cannot compute
a function of the output of a Bayesian differential private algorithm  and reduce its privacy guarantee.

\end{abstract}

%6.74997pt plus 4.0pt minus 2.0pt 6.74997pt plus 4.0pt minus 2.0pt 6.74997pt plus 4.0pt minus 2.0pt 0.0pt plus 4.0pt

% \setlength{\belowdisplayskip}{4pt plus 2pt minus 2pt}%
% \setlength{\belowdisplayshortskip}{4pt plus 2pt minus 2pt}%
%  \setlength{\abovedisplayskip}{4pt plus 2pt minus 2pt}%
% \setlength{\abovedisplayshortskip}{0.0pt plus 2.0pt}

%\the\belowdisplayskip
%
%\the\belowdisplayshortskip
%
%\the\abovedisplayskip
%
%\the\abovedisplayshortskip

%\vspace{10pt}
%
%\begin{IEEEkeywords}
%key predistribution, node degree, random graph, random intersection
%graph, random key graph, security, topological properties, wireless
%sensor networks.
% \end{IEEEkeywords}

\begin{IEEEkeywords}
Differential privacy, Bayesian differential privacy, {sequential composability}, {parallel composability}, and \vspace{-5pt} {post-processing}.
  \end{IEEEkeywords}
%, random key graph,
%isolated node \cite{yagan-heter-IT,yagan-heter-k1,yagan-heter-k}

\section{Introduction\vspace{0pt}}

Differential privacy by Dwork~\textit{et~al.} \cite{Dwork2006,dwork2006calibrating} is a robust privacy standard that has been
used in a range of data analysis tasks, since it provides a rigorous foundation for defining and
preserving privacy. Differential privacy has received considerable attention in the literature \cite{zhang2016privtree,loucost,wang2017privsuper,blocki2016differentially,KamalikaChaudhuriAllerton17,shokri2015privacy}. Apple has incorporated differential privacy into its mobile operating system iOS 10~\cite{apple}.  Google has implemented a differentially private tool called
RAPPOR in the Chrome browser to collect information about clients~\cite{erlingsson2014rappor}.
 A randomized algorithm $Y$ satisfies $\epsilon$-differential  privacy
if for any adjacent databases $x$ and $x'$ differing in one record,  and for any event $E$, it holds that
\mbox{$\pr[Y(x) \in E] \leq e^{\epsilon} \pr[Y(x') \in E],$} where~$\pr[\cdot]$ denotes the probability throughout this paper.  Intuitively, under differential privacy, an adversary
given access to the output does not have much confidence to determine
whether it was sampled from the probability distribution
generated by the algorithm when the database is $x$ or when the database is $x'$.

%\textbf{Bayesian differential privacy (BDP).}

Despite the powerfulness of differential privacy, it has recently  been observed by Kifer and Machanavajjhala \cite{kifer2011no} (see also \cite{fullpaper,chen2014correlated,he2014blowfish,kifer2012rigorous,Changchang2016,KamalikaChaudhuriSIGMOD17,zhu2015correlated}) that differential privacy may not work as expected when the data tuples  have dependencies. To extend differential privacy when data tuples have dependencies,
Yang~\textit{et~al.}  \cite{yang2015bayesian} introduce the notion of Bayesian differential privacy as follows. For a database $x$ with $n$ tuples,  let $i \in \{1,2,\ldots,n\}$ be a tuple index in the
database and~$K \subseteq \{1,2,\ldots,n\}\setminus \{i\}$ be a tuple index set. An
adversary denoted by $A(i, K) $    knows
the values of all tuples in $K$ (denoted by $x_K$)  and attempts
to attack the value of tuple $i$ (denoted by $x_i$). For a randomized   mechanism~\mbox{$Y = \pr[y \in \mathcal{Y} \con x]$}  on database $x$, the Bayesian differential privacy leakage (BDPL) of $Y$ with respect to the adversary  $A(i, K) $ is $\text{BDPL}_{A(i, K)}(Y) = \text{sup}_{x_i, x_i', x_K,  \mathcal{Y}} \ln \frac{\pr[y \in \mathcal{Y} \con x_i, x_{K}]}{\pr[y \in \mathcal{Y} \con x_i', x_K]} $. The mechanism $Y$ satisfies \mbox{$\epsilon$-Bayesian differential privacy} if $ \text{BDPL}_{A(i, K)}(Y) \leq \epsilon$ for any $A(i, K)$.

 %
%\subsection{composability and Post-Processing Properties of Bayesian Differential Privacy}\label{composabilityandpost-processing}

In this paper, we formally show that similar to differential privacy, Bayesian differential privacy has the following nice properties: \emph{sequential composability}, \mbox{\emph{parallel composability}}, and \emph{post-processing}, as detailed below \cite{Dwork2006,mcsherry2009privacy}.  The idea behind \emph{sequential composability} is that if we have $m$ algorithms~$Y_1, Y_2, \ldots, Y_m$,
where~$Y_{\ell}$
is independently $\epsilon_{\ell}$-Bayesian differential private for ${\ell}=1,2,\ldots,m$, then by feeding the
result of $Y_1$ into $Y_2$, the
result of $Y_2$ into $Y_3$, and so on, we will finally have an
$\sum_{\ell=1}^m \epsilon_{\ell}$-Bayesian differential private algorithm. For \emph{parallel composability}, we consider the situation where a   database is partitioned into $m$ disjoint subsets. The~\mbox{$\ell$-th subset}  is input to a Bayesian differential private algorithm~$Y_{\ell}$, for ${\ell}=1,2,\ldots,m$. Then the parallel composition of~\mbox{$Y_1$,  $Y_2$, $\ldots$,  $Y_m$}  will be $\max_{\ell=1}^m \epsilon_{\ell}$-Bayesian differential private. The \mbox{\emph{post-processing}} property means that a data analyst,
without additional knowledge about the private database, cannot compute
a function of the output of a Bayesian differential private algorithm  and reduce its privacy guarantee.

The rest of the paper is organized as follows. Section
\ref{sec:main:res} presents the results on the {sequential composability}, {parallel composability}, and {post-processing} properties of Bayesian \mbox{differential} privacy. We elaborate their proofs in Sections \ref{secprofco}.  Section \ref{related} surveys related work, and
Section \ref{sec:Conclusion} concludes the paper.

\section{The Results} \label{sec:main:res}

%
%\section{Proving composability and post-processing properties of Bayesian differential privacy on Page \pageref{composabilityandpost-processing}}  \label{secBDPCompositionpost-processing}
%

We prove that similar to differential privacy, the notion of Bayesian differential privacy has the following properties: \emph{sequential composability}, \emph{parallel composability}, and \mbox{\emph{post-processing}.}

\subsection{Sequential composability}

 The idea behind sequential composability is that if we have~$m$ algorithms
which are each independently Bayesian differential private, we would like to feed the
results from the first into the second, and so on, without completely sacrificing privacy. Sequential
composability allows us to do this.

\begin{thm}[\textbf{Sequential composability}] \label{thm:SequentialComposition}

Let $x$ be the database. For ${\ell}=1,2,\ldots,m$, suppose $a_{\ell}$ represents an auxiliary
input or intermediate output of an algorithm. We have $m$ algorithms $Y_{\ell}(x, a_{\ell})$ for ${\ell}=1,2,\ldots,m$. Furthermore, assume that $Y_{\ell}$
is independently $\epsilon_{\ell}$-Bayesian differential private for ${\ell}=1,2,\ldots,m$. Consider a sequence of computations \\$Y_1(x)=z_1,$\\$ Y_2(x,z_1)=z_2, $\\$\ldots, $\\$ Y_m(x,z_1,z_2,\ldots,z_{m-1})=z_m$,\\ where the expression here is general enough to cover all cases regardless whether the input of  $Y_{\ell}$ for ${\ell}=1,2,\ldots,m$ may or may not include partial or all outputs $z_1,z_2,\ldots,z_{\ell-1}$ of $Y_1,Y_2,\ldots,Y_{\ell-1}$. Let the mechanism $Y$ denote the sequential composition of $Y_1$,  $Y_2$, $\ldots$,  $Y_m$; i.e., $Y(x) = z_m$. Then $Y$ achieves $\sum_{\ell=1}^m \epsilon_{\ell}$-Bayesian differential privacy.
\end{thm}

We show Theorem \ref{thm:SequentialComposition} in Section \ref{secprofco}.

\subsection{Parallel composability}

 Now we consider the situation where a single database is partitioned
into $m$ disjoint subsets. Each subset is input to a Bayesian differential private algorithm. Then we consider the parallel composition of the $m$ algorithms. Specifically, we present the following theorem.

\begin{thm}[\textbf{Parallel composability}] \label{thm:ParallelComposition}

Let $x$ be the database whose tuples are indexed from $1$ to $n$. Let $H_1$, $H_2$, $\ldots$, $H_m$ be a partition of the index set $\{1,2,\ldots,n\}$. For ${\ell}=1,2,\ldots,m$, suppose $a_{\ell}$ represents  an auxiliary
input or intermediate output of an algorithm. We have $m$ algorithms $Y_{\ell}(x_{H_{\ell}}, a_{\ell})$ for ${\ell}=1,2,\ldots,m$. Furthermore, assume that $Y_{\ell}$
is independently $\epsilon_{\ell}$-Bayesian differential private for ${\ell}=1,2,\ldots,m$. Consider a sequence of computations \\$Y_1(x_{H_1})=y_1, $\\$Y_2(x_{H_2},y_1)=y_2, $\\$\ldots,$\\$ Y_m(x_{H_m},y_1,y_2,\ldots,y_{m-1})=y_m$,\\  where the expression here is general enough to cover all cases regardless whether the input of  $Y_{\ell}$ for ${\ell}=1,2,\ldots,m$ may or may not include partial or all outputs $y_1,y_2,\ldots,y_{\ell-1}$ of $Y_1,Y_2,\ldots,Y_{\ell-1}$. Let the mechanism $Y$ denote the parallel composition of $Y_1$,  $Y_2$, $\ldots$,  $Y_m$; i.e., $Y(x)=y_1 || y_2 || \ldots || y_m$ for $y_1 $, $ y_2 $, $ \ldots $, $ y_m$ defined above, where ``$||$'' means concatenation. Then $Y$ achieves $\max_{\ell=1}^m \epsilon_{\ell}$-Bayesian differential privacy.
\end{thm}
 Theorem \ref{thm:ParallelComposition} will be proved in the full version \cite{fullpaper2} due to space limitation.

\subsection{Post-Processing}
 Similar to differential privacy, our Bayesian differential privacy is also immune to post-processing: A data analyst,
without additional knowledge about the private database, cannot compute
a function of the output of a Bayesian differential private algorithm  and reduce its privacy guarantee. Specifically, we have the following theorem.

\begin{thm}[\textbf{Post-Processing}] \label{thm:PostProcessing} Let $Y$ be an $\epsilon$-Bayesian differential private algorithm. Let $Z$ be an arbitrary randomized mapping ($Z$ sees the output of $Y$, but not the database). Then $Z \circ Y$ is $\epsilon$-Bayesian differential private, where $Z \circ Y$ is defined such that $Z \circ Y(x) = Z\big(Y(x)\big)$ for each database $x$.
  \end{thm}

We establish Theorem \ref{thm:PostProcessing} in the full version \cite{fullpaper2} due to space limitation.

\section{Proofs} \label{secprofco}

In this section, we prove the theorems.
Without loss of generality, we consider discrete outputs so we use   probability $\bp{\cdot}$ below. If the output is continuous, we just replace probability $\bp{\cdot}$ with probability density function $\bfu{\cdot}$, and the proof follows accordingly. We introduce some notation as follows. The database under  consideration is modeled by a random variable $X=[{X}_1,{X}_2, \ldots, {X}_n]$, where ${X}_j$ for each $j\in  \{1,\iffalse 2,\fi \ldots,n\}$ is a {tuple}, which is also a {random variable}. Let database $x=[x_1,x_2, \ldots,x_n]$ be an instantiation of $X$, so that each ${x}_j$ denotes an instantiation of ${X}_j$. An
adversary denoted by $A(i, K) $    knows
the values of all tuples in $K$ (denoted by $x_K$)  and attempts
to attack the value of tuple $i$ (denoted by $x_i$). We define \mbox{$\overline{K}:=\{1,2,\ldots,n\}\setminus \{i\} \setminus K$}. Then for a randomized   mechanism~\mbox{$Y = \pr[y \in \mathcal{Y} \con x]$}  on database~$x$, we write \mbox{$\pr[y \in \mathcal{Y} \con x_i, x_{K}]$}  as \mbox{$\bigp{Y(x_i,x_K,X_{\overline{K}}) \in \mathcal{Y}}$}. For an index set $S$, we group $x_j$ (resp., $X_j$) for $j \in S$ and write $x_S$ (resp., $X_S$). Hence,
 the Bayesian differential privacy leakage (BDPL) of $Y$ with respect to the adversary  $A(i, K) $ is
 \begin{align}
& \text{BDPL}_{A(i, K)}(Y)\nonumber \\[5pt]
& \quad = \text{sup}_{x_i, x_i', x_K,  \mathcal{Y}} \ln \frac{\pr[y \in \mathcal{Y} \con x_i, x_{K}]}{\pr[y \in \mathcal{Y} \con x_i', x_K]}
\nonumber \\[5pt]
& \quad = \text{sup}_{x_i, x_i', x_K,  \mathcal{Y}} \ln \frac{\bigp{Y(x_i,x_K,X_{\overline{K}}) \in \mathcal{Y}}}{\bigp{Y(x_i',x_K,X_{\overline{K}}) \in \mathcal{Y}}}
\nonumber.
\end{align}
%  $\text{BDPL}_{A(i, K)}(Y) = \text{sup}_{x_i, x_i', x_K,  \mathcal{Y}} \ln \frac{\pr[y \in \mathcal{Y} \con x_i, x_{K}]}{\pr[y \in \mathcal{Y} \con x_i', x_K]} = \text{sup}_{x_i, x_i', x_K,  \mathcal{Y}} \ln \frac{\bigp{Y(x_i,x_K,X_{\overline{K}}) \in \mathcal{Y}}}{\bigp{Y(x_i',x_K,X_{\overline{K}}) \in \mathcal{Y}}}  $.
 The mechanism $Y$ satisfies \mbox{$\epsilon$-Bayesian differential privacy} if $$ \text{BDPL}_{A(i, K)}(Y) \leq \epsilon$$ for any $A(i, K)$, where $i \in \{1,2,\ldots,n\}$ and~$K \subseteq \{1,2,\ldots,n\}\setminus \{i\}$.

% \subsection{Proof of Theorem \ref{thm:SequentialComposition}} \label{prfthm1}

\textbf{Proof of Theorem \ref{thm:SequentialComposition}:}

We have $z_m$ as the output of algorithm $Y$. Then we consider $\bigp{Y(x_i,x_K,X_{\overline{K}})=z_m}$ so that  $i$ is the index of the tuple to be protected by the mechanism (i.e., inferred by the adversary), where \mbox{$i \in \{1,2,\ldots,n\} $,} \mbox{$K \hspace{2pt}\subseteq\hspace{2pt} \{1,\hspace{2pt}2,\hspace{2pt}\ldots,\hspace{2pt}n\}\hspace{2pt}\setminus\hspace{2pt}\{i\}$,}  \mbox{$\overline{K} \hspace{2pt}\de\hspace{2pt} \{1,\hspace{2pt}2,\hspace{2pt}\ldots,\hspace{2pt}n\}\hspace{2pt}\setminus\hspace{2pt}\{i\}\hspace{2pt}\setminus\hspace{2pt} K$,} \mbox{$x_i \in \textrm{domain}(X_i)$,} \mbox{$x_i' \in \textrm{domain}(X_i)$,} \mbox{$x_K \in \textrm{domain}(X_K)$.}  By the law of total probability, it follows that
\begin{align}
& \bigp{Y(x_i,x_K,X_{\overline{K}})=z_m} \nonumber \\[5pt] &
 = \sum_{z_1,z_2,\ldots,z_{m-1}} \nonumber \\[5pt] & \bp{ \begin{array}{l} Y_1(x_i,x_K,X_{\overline{K}})=z_1, \\[5pt] Y_2(x_i,x_K,X_{\overline{K}},z_1)=z_2, \\[5pt] \ldots, \\[5pt] Y_m(x_i,x_K,X_{\overline{K}},z_1,z_2,\ldots,z_{m-1})=z_m \end{array} }  . \label{SequentialComposition1}\end{align}

Since $Y_1$,  $Y_2$, $\ldots$,  $Y_m$ are independent, (\ref{SequentialComposition1}) further induces
\begin{align}  &  \bigp{Y(x_i,x_K,X_{\overline{K}})=z_m}
 \nonumber \\[5pt]  &  = \sum_{z_1,z_2,\ldots,z_{m-1}} \nonumber \\[5pt]  & \left\{\begin{array}{l} \bigp{Y_1(x_i,x_K,X_{\overline{K}}) = z_1} \\[5pt] \times \bigp{Y_2(x_i,x_K,X_{\overline{K}},z_1) = z_2} \\[5pt]  \times \ldots  \\[5pt] \times \bigp{Y_m(x_i,x_K,X_{\overline{K}},z_1,z_2,\ldots,z_{m-1}) = z_m}\end{array}\right\} .
 \label{SequentialComposition2}\end{align}

Now we consider the scenario where   $x_i$ in the database is replaced by $x_i'$.  Since $Y_{\ell}$
is $\epsilon_{\ell}$-Bayesian differential private for ${\ell}=1,2,\ldots,m$, we obtain for $x_i \in \textrm{domain}(X_i)$, $x_i' \in \textrm{domain}(X_i)$, $x_K \in \textrm{domain}(X_K)$ that
\begin{align}
\begin{rrcases}   &  \bigp{Y_1(x_i,x_K,X_{\overline{K}})=z_1} \\[3pt] &  \leq \exp(\epsilon_1) \times \bigp{Y_1(x_i',x_K,X_{\overline{K}})=z_1}, \\[5pt]  & \bigp{Y_2(x_i,x_K,X_{\overline{K}},z_1)=z_2} \\[3pt]  &  \leq \exp(\epsilon_2) \times  \bigp{Y_2(x_i',x_K,X_{\overline{K}},z_1)=z_2}, \\[5pt]   & \ldots, \\[5pt] & \bigp{Y_m(x_i,x_K,X_{\overline{K}},z_1,z_2,\ldots,z_{m-1})=z_m} \\[3pt] &   \leq  \exp(\epsilon_m) \times  \\[3pt] & \quad \times  \bigp{Y_m(x_i',x_K,X_{\overline{K}},z_1,z_2,\ldots,z_{m-1})=z_m}.
\end{rrcases}\label{SequentialComposition3}
\end{align}

Applying (\ref{SequentialComposition3}) to  (\ref{SequentialComposition2}), we obtain
\begin{align}   &   \bigp{Y(x_i,x_K,X_{\overline{K}})=z_m}     \nonumber \\[5pt] &  \leq  \sum_{z_1,z_2,\ldots,z_{m-1}}\nonumber \\[5pt] &  \left\{  \begin{array}{l} \exp(\epsilon_1) \cdot \bigp{Y_1(x_i',x_K,X_{\overline{K}})=z_1} \\[5pt] \times \exp(\epsilon_2) \cdot  \bigp{Y_2(x_i',x_K,X_{\overline{K}},z_1)=z_2} \\[5pt] \times \ldots  \\[5pt] \times \exp(\epsilon_m) \cdot \bigp{Y_m(x_i',x_K,X_{\overline{K}},z_{m-1})=z_m}\end{array}  \right\} \nonumber \\[5pt] &     = \exp\bigg(\sum_{\ell=1}^m \epsilon_{\ell}\bigg)\times \sum_{z_1,z_2,\ldots,z_{m-1}}\nonumber \\[5pt] &  \left\{  \begin{array}{l} \bigp{Y_1(x_i',x_K,X_{\overline{K}})=z_1} \\[5pt] \times \bigp{Y_2(x_i',x_K,X_{\overline{K}},z_1)=z_2} \\[5pt] \times \ldots  \\[5pt] \times \bigp{Y_m(x_i',x_K,X_{\overline{K}},z_{m-1})=z_m}\end{array}  \right\} .\label{SequentialComposition4}\end{align}

Replacing $x_i$ by $x_i'$ in (\ref{SequentialComposition1}), we get
\begin{align}
  &  \bigp{Y(x_i',x_K,X_{\overline{K}})=z_m}
   \nonumber \\ & = \sum_{z_1,z_2,\ldots,z_{m-1}}\nonumber \\ &  \left\{\begin{array}{l} \bigp{Y_1(x_i',x_K,X_{\overline{K}})=z_1} \\[5pt]  \times \bigp{Y_2(x_i',x_K,X_{\overline{K}},z_1)=z_2} \\[5pt]  \times \ldots  \\[5pt]  \times \bigp{Y_m(x_i',x_K,X_{\overline{K}},z_{m-1})=z_m}\end{array}\right\}. \label{SequentialComposition5}
\end{align}

Finally, (\ref{SequentialComposition4}) and (\ref{SequentialComposition5}) together imply the desired result
\begin{align}  &  \bigp{Y(x_i,x_K,X_{\overline{K}})=z_m}  \nonumber \\ &  \leq \exp\bigg(\sum_{\ell=1}^m \epsilon_{\ell}\bigg)\cdot \bigp{Y(x_i',x_K,X_{\overline{K}})=z_m}.   \label{SequentialComposition5fl}
\end{align}
We can show (\ref{SequentialComposition5fl}) for any $i \in \{1,2,\ldots,n\}$, $K \subseteq \{1,2,\ldots,n\}\setminus \{i\}$, $x_i \in \textrm{domain}(X_i)$, $x_i' \in \textrm{domain}(X_i)$, $x_K \in \textrm{domain}(X_K)$, $z_m\in \textrm{range}(Y)$.
Hence, the mechanism $Y$ is $\sum_{\ell=1}^m \epsilon_{\ell}$-Bayesian differential private. %\hfill\fsquare

\textbf{Proofs of Theorems \ref{thm:ParallelComposition} and \ref{thm:PostProcessing}:} Due to space limitation, we establish Theorems \ref{thm:ParallelComposition} and \ref{thm:PostProcessing} in the full version~\cite{fullpaper2}.

\section{Related Work}  \label{related}

 The notion of {differential privacy} \cite{Dwork2006,dwork2006calibrating}  provides a rigorous foundation for privacy protection. Intuitively, differential privacy implies that changing one entry in the
database does not significantly change the query output, so that an adversary, seeing the query output and even knowing all records except the one to be inferred,  draws almost the same conclusion on whether or not a record is in the database. Differential privacy is shown to satisfy the properties of sequential composability, parallel composability, and post-processing
\cite{Dwork2006,mcsherry2009privacy}. Recently, Kairouz~\textit{et~al.}~\cite{KairouzIT2017} investigate the overall privacy  cost for the composition of differential private algorithms. Their result improves those in prior work \cite{Dwork2006,dwork2006calibrating,dwork2009differential,dwork2010boosting}.

Although differential privacy has received considerable interest in the literature \cite{qin2016heavy,kasiviswanathan2014semantics,xiao2015protecting,Jiang:2013:PTD:2484838.2484846,DDPWPES,zhang2017cost,winograd2017framework,bun2017make,lyu2017understanding,yao2017privacy,han2017differentially}, it has   been observed by Kifer and Machanavajjhala \cite{kifer2011no} (see also \cite{fullpaper,chen2014correlated,he2014blowfish,kifer2012rigorous,Changchang2016,KamalikaChaudhuriSIGMOD17,zhu2015correlated}) that differential privacy may not work as expected when the data tuples  have dependencies.
To extend differential privacy for correlated data, prior studies have investigated
various privacy metrics \cite{chen2014correlated,he2014blowfish,kifer2012rigorous,Changchang2016,KamalikaChaudhuriSIGMOD17,DDPAllerton,zhu2015correlated}. One of the metrics receiving much attention is the notion of Bayesian differential privacy  introduced by Yang~\textit{et~al.}~ \cite{yang2015bayesian}. Yang~\textit{et~al.}~\cite{yang2015bayesian} further present a mechanism that is only for the sum query on a Gaussian Markov random field with positive correlations and its extension to a discrete domain.
In contrast, Zhao~\textit{et~al.}~\cite{fullpaper} propose mechanisms to achieve Bayesian differential privacy for any query on databases with arbitrary tuple correlations.

 Kifer and Machanavajjhala \cite{kifer2012rigorous} generalize differential privacy to the Pufferfish framework, which takes into consideration the generation of the database and the adversarial belief about the database.  Li~\textit{et~al.}~\cite{Li-CCS2013} propose membership privacy in  consideration of the adversary's prior beliefs as well.   He~\textit{et~al.}~\cite{he2014blowfish} study a subclass of
the Pufferfish framework, named the Blowfish framework, which  uses deterministic
policy constraints instead of probabilistic correlations to specify adversarial
knowledge about the database. Very recently,
Song~\textit{et~al.}~\cite{KamalikaChaudhuriSIGMOD17} propose a general mechanism to achieve Pufferfish privacy. Song and Chaudhuri~\cite{KamalikaChaudhuriAllerton17} further show composition Properties of Pufferfish privacy for time-series data. Zhu~\textit{et~al.}~\cite{zhu2015correlated} leverage   linear relationships among tuples, but this approach does not
satisfy any rigorous privacy metric.  Liu~\textit{et~al.}~\cite{Changchang2016} present a Laplace mechanism that
 handles pairwise correlations. Xiao and Xiong~\cite{xiao2015protecting} address differential privacy under temporal correlations in the context of location privacy.
 Kasiviswanathan and Smith \cite{kasiviswanathan2014semantics}
 introduce  the notion of semantic privacy by modeling the external knowledge
 via a prior probability distribution, and modeling  conclusions  via the corresponding posterior distribution.

Dwork and Rothblum~\cite{dwork2016concentrated} recently proposed the notion of {concentrated differential privacy}, a relaxation of differential privacy enjoying
better accuracy than     differential privacy without
compromising on cumulative privacy loss over multiple computations. Motivated by~\cite{dwork2016concentrated}, Bun and Steinke~\cite{bun2016concentrated} suggest a relaxation of
concentrated differential privacy.  Jorgensen~\textit{et~al.}~\cite{jorgensen2015conservative} introduce  a new privacy definition called personalized differential
privacy, a generalization of differential privacy in which
users specify a personal privacy requirement for their data.
%They present a mechanism for achieving personalized differential privacy, inspired by the well-known exponential mechanism of differential privacy.

%Mironov \mbox{\textit{et al.}}~\cite{mironov2009computational} present several relaxations of differential privacy which require privacy guarantees to hold only against  computationally bounded adversaries.

\section{Conclusion}
\label{sec:Conclusion}

Bayesian differential privacy has been recently proposed to broaden the application scenarios of differential privacy when data records have dependencies.  In this paper, we formally show that Bayesian differential privacy   preserves three nice properties of differential privacy: sequential composability,  \mbox{parallel composability}, and post-processing.

\iffalse

The idea behind \emph{sequential composability} is that if we have $m$ algorithms $Y_1, Y_2, \ldots, Y_m$,
where $Y_{\ell}$
is independently $\epsilon_{\ell}$-Bayesian differential private for ${\ell}=1,2,\ldots,m$, then by feeding the
result of $Y_1$ into $Y_2$, the
result of $Y_2$ into $Y_3$, and so on, we will finally have an
$\sum_{\ell=1}^m \epsilon_{\ell}$-Bayesian differential private algorithm. For \emph{parallel composability}, we consider the situation where a   database is partitioned into $m$ disjoint subsets. The $\ell$-th subset is input to a Bayesian differential private algorithm $Y_{\ell}$, for ${\ell}=1,2,\ldots,m$. Then the parallel composition of $Y_1$,  $Y_2$, $\ldots$,  $Y_m$ will be $\max_{\ell=1}^m \epsilon_{\ell}$-Bayesian differential private. The \emph{post-processing} property means that a data analyst,
without additional knowledge about the private database, cannot compute
a function of the output of a Bayesian differential private algorithm  and reduce its privacy guarantee.

 \fi

  \let\OLDthebibliography\thebibliography
 \renewcommand\thebibliography[1]{
   \OLDthebibliography{#1}
   \setlength{\parskip}{0pt}
   \setlength{\itemsep}{1pt plus 0ex}
 }

% \linespread{1.15}

\end{document}